\documentclass[a4paper,11pt]{article}
\usepackage{jheppub}
\usepackage{amssymb}   
\usepackage{amsmath}
\usepackage{mathtools}
\usepackage{makeidx}
\usepackage{graphicx}
\usepackage{axodraw4j}
\usepackage{caption}
\usepackage{subcaption}
\usepackage[utf8]{inputenc}
\usepackage{feynmp-auto}
\usepackage{graphicx}
\usepackage[draft]{fixme}

\def\beq{\begin{equation}}
\def\eeq{\end{equation}}
\def\bea{\begin{eqnarray}}
\def\eea{\end{eqnarray}}

\def\f21{{}_2F_{1}}
\def\eps{\epsilon}

\hypersetup{colorlinks = false}

\def\bsp#1\esp{\begin{split}#1\end{split}}

\title{
Four-graviton scattering to three loops in ${\mathcal N}=8$ supergravity
\bigskip
\bigskip}

\preprint{MIT-CTP/5103, MPP-2019-25}

\author{J.\ M.\ Henn$^{a}$, B.\ Mistlberger$^{b}$}

\affiliation{
$^a$ Max-Planck-Institut f{\"u}r Physik, Werner-Heisenberg-Institut, 80805 M{\"u}nchen, Germany\\
$^b$ Center for Theoretical Physics, Massachusetts Institute of Technology, Cambridge, MA 02139, USA}

\emailAdd{henn@mpp.mpg.de}
\emailAdd{bernhard.mistlberger@gmail.com}

\abstract{
We compute the three-loop scattering amplitude of four gravitons in ${\mathcal N}=8$ supergravity. 
Our results are analytic formulae for a Laurent expansion of the amplitude in the regulator of dimensional regularisation. 
The coefficients of this series are closed formulae in terms of well-established harmonic poly-logarithms. 
Our results display a remarkable degree of simplicity and represent an important stepping stone in the exploration of the structure of scattering amplitudes. 
In particular, we observe that to this loop order the four graviton amplitude is given by uniform weight $2L$ functions, where $L$ is the loop order.
 }
\keywords{N=8 supergravity, scattering amplitudes}

\begin{document}
\notoc
\maketitle
\newpage

\tableofcontents

\newpage 

\section{Introduction}
\label{sec:intro}

Scattering amplitudes in gravity theories, and in particular in 
$\mathcal{N}=8$ supergravity, are interesting for a variety of reasons.
Many studies are devoted to the question whether $\mathcal{N}=8$ supergravity 
is ultraviolet (UV) finite perturbatively.
The theory has been explicitly shown to be UV finite in four dimensions up to four loops \cite{Bern:2009kd},
and arguments based on counterterms do not expect a divergence before seven loops \cite{Bossard:2011tq}. 
This order is currently out of reach of perturbative calculations, see \cite{Bern:2018jmv} and references
therein for the state of the art. 

One may hope to make progress on this question from another angle. 
Amplitudes in supergravity are closely related to their Yang-Mills counterparts.
At tree-level, the relationship comes in the form of KLT \cite{Kawai:1985xq,Bern:1998ug} and the related
BCJ color-kinematics duality and double-copy construction \cite{Bern:2010ue}. 
The double-copy relationship is expected to hold at loop level,
where it reduces the problem of constructing gravity integrands to that of 
calculating much simpler gauge-theory ones. 
In this sense, there is a relationship between $\mathcal{N}=8$ supergravity and
$\mathcal{N}=4$ super Yang-Mills (sYM), a finite quantum field theory. 
This begs the question whether a concise relationship exists for 
the {\it integrated} amplitudes. This is far from obvious, as the double copy
procedure in general changes power counting and other properties 
of individual loop integrands.

The BCJ relations have also been applied at the level of classical gravity 
solutions \cite{Luna:2016hge,Luna:2016due,Goldberger:2017vcg}, and recent work \cite{Luna:2017dtq,Kosower:2018adc,Bern:2019nnu} applies amplitude methods to compute classical quantities in gravity~\cite{Antonelli:2019ytb}.

Gravity amplitudes have also received attention in the context for renewed interest in soft theorems. See ref. \cite{Strominger:2017zoo} for a review of connections between soft theorems, asymptotic symmetries, and the memory effect of gravitational radiation.
New subleading soft theorems have been shown to hold at the level of tree-level amplitudes and loop integrands \cite{Cachazo:2014fwa}, while at loop level corrections are expected \cite{Bern:2014oka,He:2014bga,DiVecchia:2018dob,Larkoski:2014bxa}.

In this paper we focus on amplitudes in $\mathcal{N}=8$ supergravity, and their relations to $\mathcal{N}=4$ super Yang-Mills.
We are motivated by the many special properties found in the latter theory, and find it interesting to look for hints of simplicity in the former theory.
The key properties we focus on are the structure of infrared divergences, hidden symmetries, and the analytic structure of loop corrections.

The infrared (IR) properties of gravity are much simpler compared to gauge theories. Gravity amplitudes are free from collinear divergences, and soft divergences can be shown to exponentiate in a way similar to Abelian theories. As a result, the amplitudes are much less singular in the infrared compared to their Yang-Mills counterparts. One may easily take into account the exponentiated divergences and define an IR-finite remainder function. 
It is worth mentioning that these simple IR properties typically become manifest only upon summing over all permutations of the contributing planar and non-planar Feynman diagrams.
We take this as an encouraging sign that certain simple properties of (super)gravity theories may be somewhat hidden at the level of the loop integrand.

Planar $\mathcal{N}=4$ super Yang-Mills has a hidden dual (super)conformal \cite{Drummond:2006rz,Alday:2007hr,Drummond:2007au,Drummond:2008vq,Berkovits:2008ic,Brandhuber:2008pf} or Yangian symmetry \cite{Drummond:2009fd}. This hidden symmetry is best thought of as a generalization of the well-known Laplace-Runge-Lenz symmetry of the hydrogen atom in quantum mechanics \cite{Caron-Huot:2014gia}. It is an exact symmetry of bound states in $\mathcal{N}=4$ super Yang-Mills.
The symmetry has far-reaching consequences. For example, it entirely fixes the functional form of four- and five-particle amplitudes, and severely restricts higher-point amplitudes \cite{Drummond:2007au}.
First hints for a possible generalization of dual conformal symmetry to non-planar scattering amplitudes appeared in refs. \cite{Bern:2017gdk,Ben-Israel:2018ckc,Bern:2018oao,Chicherin:2018wes} and for form factors in \cite{Bianchi:2018rrj}. 
At present, the version of the symmetry discussed in \cite{Bern:2017gdk,Bern:2018oao,Chicherin:2018wes} applies to certain individual integrals only (as opposed to the full amplitude), while the version of \cite{Ben-Israel:2018ckc} acts on objects related to, but slightly different from, the original amplitudes.
See also \cite{Caron-Huot:2018ape} for a discussion of hidden symmetries at the level of bound states of black holes in $\mathcal{N}=8$ supergravity.

An empirically observed property of scattering amplitudes in $\mathcal{N}=4$ super Yang-Mills has to do with the properties of the functions appearing in them. In all cases where the answer is given by multiple polylogarithms, it has been observed that the transcendental weight of the expressions (i.e. the number of integrations) at $L$ loops is always $2L$. We refer to this as the maximal weight property of $\mathcal{N}=4$ super Yang-Mills. This is closely related to observations that for certain anomalous dimensions, the $\mathcal{N}=4$ sYM answer may be obtained from the corresponding QCD results by retaining the maximal weight piece only \cite{Kotikov:2004er,KOTIKOV2007217}.

It is by now rather well understood which Feynman integrals evaluate to maximal weight integrals.
Conjecturally, Feynman integrals whose integrands can be written as {\it dlog} forms have this property \cite{ArkaniHamed:2010gh,Henn:2013pwa,Arkani-Hamed:2014via,WasserMSc}.
Note that this implies the absence of double (or higher poles) at the level of the loop integrand.
This conjecture can be motivated by, and is closely connected to the differential equations that
such functions satisfy \cite{Henn:2013pwa,Frellesvig:2017aai,Harley:2017qut,Primo:2016ebd}.
In the literature, the {\it dlog} property is often discussed for four-dimensional loop integrands.
In general, this is insufficient, as integrands vanishing in four dimensions may influence the 
weight properties of the integrated answer. See \cite{Chicherin:2018old} for a recent proposal on
how to analyze such terms.

Given the above remarks, it is important to have access to integrated amplitudes in order to ultimately decide whether
amplitudes are given by maximal weight functions.
The two-loop five-graviton amplitude in $\mathcal{N}=8$ supergravity \cite{Chicherin:2019xeg,Abreu:2019rpt} has maximal weight.
The one- and two-loop four-graviton amplitudes share the same property \cite{Naculich:2008ew,Brandhuber:2008tf}. To appreciate that these are remarkable properties, we emphasize that amplitudes with less than maximal supersymmetries are known to contain terms with lower weight \cite{BoucherVeronneau:2011qv}.
This happens even in the case of UV-finite theories.

The integrand for three-loop four-graviton amplitudes in $\mathcal{N}=8$ supergravity is available in the literature in several forms that make different properties
manifest. It was first obtained in ref. \cite{Bern:2007hh} using the $D$-dimensional generalized unitarity method.
These expressions contain up to quartic terms in loop momenta that correspond to two irreducible numerators.
In this paper, the ultraviolet finiteness at three loops was explicitly shown.
In ref. \cite{Bern:2008pv}, a different expression for the integrand was given that has manifest UV power counting.
In ref. \cite{Bern:2010ue}, a form of the integrand was given that is compatible with double copy relations \cite{Bern:2008qj}.
To the best of our knowledge, an explicit {\it dlog} form of the integrand (even four-dimensional) is not available at present.
On the other hand, in section 8 of \cite{Bern:2014kca} an argument for the existence of such a representation is given. The argument goes as follows.
Suppose that for the  $\mathcal{N}=4$ sYM integrand both a BCJ representation as well as a {\it dlog} representation exist.
Note that the numerators in the {\it dlog} part are constructed such that, upon taking residues, no double poles occur.
Then one can use these numerators to construct, via BCJ, the (four-dimensional part of the) supergravity integrand. By construction, thanks
to the numerator of one of the two integrands entering the double copy, no double poles can occur at finite locations of the loop momenta.
In the same paper, it was argued that, thanks to the linearity in momentum of the BCJ numerators in the three-loop four-point amplitude,
poles at infinity are also absent at that order. The same is not necessarily true for more general amplitudes \cite{Bourjaily:2018omh}.
Therefore, conjecturally, one may expect the three-loop four-graviton amplitude in $\mathcal{N}=8$ supergravity to have uniform weight.
One of the goals of the present paper is to verify this.

In this paper we compute for the first time, and in fully analytical form, the three-loop four-graviton amplitude in $\mathcal{N}=8$ supergravity.
Our computation builds on our previous work in $\mathcal{N}=4$ sYM \cite{Henn:2016jdu}, but represents a leap in complexity of the calculation.
Our starting point is a well known formula for the integrand of the desired scattering amplitude~\cite{Bern:2007hh,Bern:2008pv,Bern:2010ue}. 
We then use cutting edge techniques for the computation of multi-loop Feynman integrals to perform the loop integration. 
This allows us to express the final answer in terms of a class of well known analytic functions - so-called harmonic poly logarithms~\cite{Remiddi:1999ew}.
We verify that our result displays the expected structure of infrared divergences and is compatible with analytic predictions for the high energy limit of the amplitude~\cite{Bartels:2012ra}.
Our result displays remarkable simplicity on which we elaborate below.
The obtained formulae will serve as key ingredients for the exploration of properties of scattering amplitudes in the future.

This paper is organized as follows.
In section \ref{sec:calculation}, we set up the notation and review the one-loop four-graviton amplitudes.
Section \ref{sec:infrared} is devoted to the infrared properties of gravity.
In section \ref{sec:analytic}, we explain in detail our procedure for analytic continuation between different physical regions.
We present our novel three-loop results in section \ref{sec:threeloops}.
Section \ref{sec:regge}, we study the soft and Regge limit of the amplitude.
We conclude in section \ref{sec:conclusion}.

\section{Notations and four-graviton amplitude at one loop}
\label{sec:calculation}

In this section we briefly discuss the overall setup and introduce some definitions.
We write the amplitude for scattering of gravitons in $\mathcal{N}=8$ super gravity in terms of 
\beq
\mathcal{A}=\mathcal{A}_0\left(1+\sum\limits_{i=1}^\infty\alpha_G \mathcal{A}^{(n)}\right)\,.
\eeq
Here $\mathcal{A}_0$ is the tree-level amplitude written in superspace \cite{ArkaniHamed:2008gz}. This has the advantage that it is completely permutation symmetric.
When specified to gravitons of helicity $--++$, for example, it becomes (see for example ref.~\cite{Bern:1998ug})
\begin{align}
\mathcal{A}_0 (--++)=i8 \pi G_N \frac{s t}{u} \left[\frac{\langle 1 2\rangle^3}{\langle 23\rangle\langle 34 \rangle\langle 41\rangle}\right]^2 \,.
\end{align}
The scattering amplitude is a function of the Lorentz invariant scalar products of the external momenta. 
We introduced the variables
\beq
s=(p_1+p_2)^2,\hspace{1cm}
t=(p_2+p_3)^2,\hspace{1cm}
u=(p_1+p_3)^2.
\eeq
Throughout this work we work regulate ultra-violet and infrared divergences of the Feynman integrals appearing in the amplitudes by taking the space time dimension to be $d=4-2\epsilon$. (The three-loop supergravity amplitude is free of ultraviolet divergences.)
The coupling constant $\alpha_G$ is related to Newton's constant $G_N$ by
\beq
\alpha_G=\frac{G_N(4\pi)^{\epsilon}}{8\pi^2 i} c_\Gamma.
\eeq
Here, we chose to absorb several factors depending on the dimensional regulator into the definition of our expansion parameter, for convenience.
In the above equation 
\beq
c_\Gamma=\frac{\Gamma(1-\epsilon)^2\Gamma(1+\epsilon)}{\Gamma(1-2\epsilon)}.
\eeq
For example, the one-loop amplitude can be written to all orders in the dimensional regulator as 
\bea
\label{eq:1loopfull}
\mathcal{A}^{(1)}&=&\frac{1}{\epsilon^2} \Bigg\{\nonumber\\
&+&s^{-\epsilon}\left[u \, _2F_1\left(-\epsilon ,-\epsilon ;1-\epsilon ;1+\frac{s}{t}\right)+t \, _2F_1\left(-\epsilon ,-\epsilon ;1-\epsilon ;1+\frac{s}{u}\right)\right]\nonumber\\
&+&t^{-\epsilon}\left[u \, _2F_1\left(-\epsilon ,-\epsilon ;1-\epsilon ;1+\frac{t}{s}\right)+s \, _2F_1\left(-\epsilon ,-\epsilon ;1-\epsilon ;1+\frac{t}{u}\right)\right]\nonumber\\
&+&u^{-\epsilon}\left[t \, _2F_1\left(-\epsilon ,-\epsilon ;1-\epsilon ;1+\frac{u}{s}\right)+s \, _2F_1\left(-\epsilon ,-\epsilon ;1-\epsilon ;1+\frac{u}{t}\right)\right]\Bigg\}.\nonumber\\
\eea
The Gauss hypergeometric functions are pure functions when expanded in the dimensional regulator.
The deepest pole in $\epsilon$ cancels among the different contributions. 
This can be seen easily as for $\epsilon=0$ the Gauss hypergeometric functions become identical to one and the term in the large round bracket just reads $2(s+t+u)=0$. Notice, that the term in the square bracket multiplying $s^{-\epsilon}$ is symmetric under the exchange of $t$ and $u$.

We define the ratio 
\beq
x=\frac{t}{s}.
\eeq
Throughout our computation we always apply momentum conservation and replace $u=-s-t$.
Furthermore, we choose $s=-1$ as the functional dependence on $s$ can easily be recreated by dimensional analysis. 
We also assume that $x\in [0,1]$ which implies that $s\leq t \leq 0$. 
This so-defined region corresponds to scattering kinematics of the gravitons with momenta $p_1$ and $p_3$ scattering into gravitons with momenta $p_2$ and $p_4$.
We will discuss the analytic continuation of this region to other regions below.
We will express our result for amplitudes at different orders in the coupling expanded in the dimensional regulator in terms of harmonic poly-logarithms (HPLs)~\cite{Remiddi:1999ew}.
HPLs are defined by
\beq
\label{eq:HPLDef}
H(a_n,\dots,a_1,x)=(-1)^{\frac{1}{2}(|a_n|+a_n)}\int_0^x \frac{dx^\prime}{x^\prime-a_n}H(a_{n-1},\dots,a_1,x^\prime),\hspace{1cm}a_i\in\{0,-1,1\}.
\eeq
We refer to $x$ as the argument of a HPL and to the $a_i$ as its indices. 
If the right-most indices are zero the HPL would be divergent and we work with the regulated definition
\beq
\label{eq:reglogdef}
H(0,\dots,0,x)=\frac{1}{n!}\log^n(x).
\eeq
The HPLs are widely used in the literature and their properties are well understood, see for example refs.~\cite{Goncharov:1998kja,Panzer:2014gra,Duhr:2011zq,Panzer:2014caa,Duhr:2012fh,Maitre:2005uu}.
Furthermore, we find it convenient to introduce the short-hand notation
\beq
H(\underbrace{0,\dots,0,\pm1}_{m},\dots,\underbrace{0,\dots,0,\pm 1}_{n},-x)=H_{\pm m,\dots,\pm n}.
\eeq

\section{Structure of infrared divergences in $\mathcal{N}=8$ supergravity}
\label{sec:infrared}
Similar to massless gauge theory scattering amplitudes graviton scattering amplitudes suffer from infrared divergences.
These divergences originate from regions where loop momenta become very low energetic (soft) or collinear to one of the external particles. 
In massless gauge theory one typically finds one double pole per loop order originating from simultaneous soft and collinear singularities.
However, it was already noted in ref.~\cite{Weinberg:1965nx} that collinear divergences cancel in physical graviton amplitudes in 
the eikonal (soft) limit.
Consequently, at most a single pole is allowed per loop order. 
We observed this cancellation already above in the case of the one-loop amplitude when expanding in the dimensional regulator and applying the kinematic constraint $s+t+u=0$. 
In fact, collinear singularities cancel for graviton scattering amplitudes, as was proven in ref.~\cite{Akhoury:2011kq} and later in SCET in ref.~\cite{Beneke:2012xa}.
Indeed, already classical gravitational radiation in the forward direction of the emitter is suppressed compared to for example electro-magnetic radiation~\cite{VanNieuwenhuizen:1973qf}.

Soft singularities in graviton scattering amplitudes were analysed in refs.~\cite{Naculich:2008ew,Naculich:2011ry,Akhoury:2011kq,White:2011yy,Beneke:2012xa}.
Modern techniques from the analysis the infrared structure of gauge theory amplitudes were employed to study the exponentiation of soft singularities. 
The infrared structure of gravity is reminiscent of the one of QED, rather than of the much more intricate structure observed in non-abelian gauge theory~\cite{Almelid:2015jia,Almelid:2017qju}. 
In fact, the soft structure of pure graviton scattering amplitudes is even simpler than the one of QED: To any loop order soft divergences in graviton scattering are given by an exponentiation of the one-loop ones,
\beq
\label{eq:softexp}
\mathcal{A}=\mathcal{A}_0e^{\alpha_G \mathcal{A}^{(1)}} \mathcal{F}.
\eeq
Here, the function $\mathcal{F}$ is free of infra-red or collinear divergences. It starts at the two loop order.
This property is particularly remarkable given the fact that we can obtain graviton amplitude integrands from the double copy of YM theory integrands.
The source of this particular simplicity is the structure of the self-coupling of gravitons. 
While in YM the self-coupling is proportional to the charge of the emitting particle, the equivalent of the charge in gravity is the four momentum of the emitting particle. 
This means that the rate for soft gravitons emitting further soft gravitons is proportional to a soft momentum and hence vanishing (see ref~\cite{White:2015wha} for a nice review).
As a result only diagrams with soft gravitons radiated from external (hard) gravitons are non-vanishing in the soft loop momentum region. 
The abelian exponentiation of such diagrams is by now well understood and leads to eq.~\eqref{eq:softexp}.
Alternatively, we may also write
\begin{align}\label{sugra_infraed}
\log 
 \frac{\mathcal{A}}{\mathcal{A}_0} 
  = \alpha_G \mathcal{A}^{(1)}+ \log \mathcal F.
\end{align}
Furthermore, we may write a perturbative expansion of the infrared finite part as
\beq
\label{eq:logamp}
\log\mathcal{F}=\sum\limits_{i=2}^\infty a_G^i F^{(i)}+\mathcal{O}(\epsilon_{IR}).
\eeq
Here, the subscript $IR$ on the dimension regulator implies that the coefficients $F^{(n)}$ may well still be ultra-violet divergent.
The finite term, sometimes called remainder, is known at two loops for $\mathcal{N}=8$ \cite{Naculich:2008ew,Brandhuber:2008tf} and for 
$\mathcal{N}\ge 4$ supergravity \cite{BoucherVeronneau:2011qv}.

Ultraviolet divergences in quantum gravity theories have received a lot of interest, culminating in a recent computation of the ultra-violet limit of $\mathcal{N}=8$ super-gravity scattering amplitudes up to five loops in ref.~\cite{Bern:2018jmv}. 
Moreover, it is current consensus that $\mathcal{N}=8$ super-gravity is indeed UV finite up to seven loops (see for example refs.\cite{Bern:2017rjw,Bern:2017tuc,Bern:2018jmv,Bern:2013uka,Carrasco:2013ypa,Freedman:2017zgq,Kallosh:2016xnm}).
Consequently, the three loop result $F^{(3)}$ obtained in this article has to be free of poles in the dimensional regulator - and indeed it is.

\section{Analytic continuation to different physical regions}
\label{sec:analytic}

Already in eq.~\eqref{eq:1loopfull} we presented results for the one-loop amplitude.
Expanded in the dimensional regulator the result becomes
\bea
\mathcal{A}^{(1)}&=&
\frac{s^{-\epsilon}}{\epsilon^2}\left[t+u +\epsilon ^2 \left(u  \text{Li}_2\left(1+\frac{s}{t}\right)+t  \text{Li}_2\left(1+\frac{s}{u}\right) \right)\right] + \text{permutations}+\mathcal{O}(\epsilon) \nonumber\\
&=&\frac{1}{\epsilon}\left[(1+x) H(-1,x)-x H(0,x)-i \pi  (1+x)\right]\nonumber\\
&&+i \pi  H(0,x)-H(-1,0,x)-H(0,-1,x)+\mathcal{O}(\epsilon),
\eea
Notice, that deriving the second line from the first line in the above equation is not entirely trivial. 
The complication lies in the fact that our amplitude has non trivial imaginary parts in the particular scattering region we chose. 
We therefore want to address at this point the problem of analytic continuation in some detail. 

The imaginary part of scattering amplitudes is defined by the external kinematics and the Feynman prescription. The latter introduces a small imaginary part in each propagator $\Delta_F(p^2)$, thereby clarifying which contours we are integrating over when performing loop integrations.
\beq
\Delta_F(p^2)=\frac{iP^{\mu\nu\rho\sigma}}{p^2-i \delta}.
\eeq
Here, $\delta$ is infinitesimal and the exact form of the real numerator $P^{\mu\nu\rho\sigma}$ is irrelevant for the current discussion. 
To understand the problem of analytic continuation of Feynman integrals better we start by writing down schematically the Feynman parameter representation of an $L$-loop four-point scalar Feynman integral,
\beq
I(s,t,u)= C \int \prod\limits_{i=1}^P dx_i x^{\nu_i-1} U(\vec{x})^{\phi_U} F(\vec{x},s,t,u)^{\phi_F}.
\eeq
Here $C$ is some function of the external variables and the dimension; $P$ is the number of propagators in the integrals, the exponents $\nu_i$ correspond to the power of the $i^{th}$ propagator and $U$ and $F$ are the first and second graph polynomials with their characteristic exponents $\phi_U$ and $\phi_F$. The particularities of the above definitions are not really important here. 
While $U$ is a positive semi-definite polynomial, $F$ is positive semi-definite only in the so-called Euclidean region defined by $s<0$, $t<0$ and $u<0$. 
In particular, we can show that for all Feynman integrals required for our computation can be put into the form
\beq
\label{eq:FPoly}
F(\vec{x},s,t,u)=-s f_1(\vec{x})-t f_2(\vec{x})-u f_3(\vec{x})+i \delta\,,
\eeq
where the $f_i(\vec{x})$ are positive semi-definite functions of the Feynman parameters. 
The above formula may be simpler in certain cases. 
For example, planar graphs will always only depend on Mandelstam invariants of consecutive momenta. 
Depending on the ordering of the external momenta of a planar Feynman integral, this could mean that the invariant u can never appear.
The case of form factor type integrals is particularly simple since they only depend on a single invariant.

If at least one of the Mandelstam variables in eq.~\eqref{eq:FPoly} is positive our F polynomial may develop a branch cut and we would have to analytically continue before we can integrate our Feynman parameters. 
Since we are imposing momentum conservation $s+t+u=0$ we are clearly working in such a region.
In order to understand how our Feynman integrals can be analytically continued it is helpful to think of all Mandelstam invariants as independent as was explored for example in refs.~\cite{Tausk:1999vh,Henn:2013nsa}.
First, one computes the desired Feynman integral in the Euclidean region and absorbs the infinitesimal imaginary part in the Mandelstam variables
\bea
s&\rightarrow &s-i\delta,\nonumber\\
t&\rightarrow &t-i\delta,\nonumber\\
u&\rightarrow &u-i\delta.
\eea
Once we expressed the desired integral in terms of a function basis we can then analytically continue to any scattering region.
For example, if we encounter a logarithm
\bea
\log\left(\frac{u-i\delta}{s-i\delta}\right)&=&\log\left(\frac{u}{s}-i\delta\frac{(s-u)}{s^2}\right)=\log\left(-\frac{u}{s}\right)-i\pi+\mathcal{O}(\delta)\nonumber\\
&=&\log(1+x)-i\pi+\mathcal{O}(\delta).
\eea
Here, we repeatedly used the fact that $\delta$ is infinitesimal. 
We find that keeping all three invariants separate is impractical.
However, the above considerations are helpful to understand how integrals with permuted external legs can be obtained and different scattering regions may be related.

Let us consider for example the case in which momentum conservation was applied and we are handed a logarithm $\log(1+x)$ without any additional information.
Next, we want to perform an exchange of external momenta $p_1$ and $p_2$ ($u\leftrightarrow t$).
We may now interpret our logarithm as a function of our Mandelstam variables as
\beq
\log(1+x)=\log\left(-\frac{u-i\delta}{s-i\delta}\right)\rightarrow \log\left(-\frac{t-i\delta}{s-i\delta}\right)=\log\left(-x+i\delta \frac{1-x}{s} \right).
\eeq
Alternatively, we may equally think that this is simply a logarithm of the ration of $t / s$.
\beq
\log(1+x)=\log\left(1+\frac{t-i\delta}{s-i\delta}\right)\rightarrow\log\left(1+\frac{u-i\delta}{s-i\delta}\right)=\log\left(-x-i \delta\frac{2+x}{s}\right).
\eeq
The imaginary part of the two logarithms considered in the above example is clearly different and leads therefore to two different results. 
 The information we were given so far is simply not enough to uniquely determine the outcome of such a permutation.

Keeping in mind that  $x\in [0,1]$, there are certain transformations, where this ambiguity does not occur.
For example,
the exchange of momenta $p_1$ and $p_3$ relates the logarithms in our problem by
\bea
x&\rightarrow &\frac{1}{x}\nonumber,\\
\log\left(x\right)&\rightarrow& -\log\left(x\right),\nonumber\\
\log\left(1+x\right)&\rightarrow& \log\left(1+x\right)-\log\left(x\right),
\eea
irrespective of whether $x=t/s$ or $x=-1-u/s$. The exchanged invariants $s$ and $t$ are both negative and no branch cut is crossed.

On the other hand, let  us consider the permutation where we exchange $p_1$ and $p_2$.
It corresponds to the transformation $x\to -1-x$, which is problematic due to the branch cuts of $\log(1+x)$ and $\log(x)$. 
Since both cuts are crossed by applying this transformation we need to ensure we know how to interpret the argument of both logarithms. 
The physical branch cuts of our amplitude occur only at points where Mandelstam invariants vanish. 
However, given that our amplitudes are comprised of HPLs only one branch cut can be made manifest in terms of explicit logarithms at any given time.
In particular this is achieved by shuffle identities of the HPL's such as
\beq
H(a_n,\dots,a_2,0,x)=H(a_n,\dots,a_2,x)\log(x)-H(0,a_n,\dots,a_2,x)-\dots-H(a_n,\dots,0,a_2,x)
\eeq
or generalisations thereof, to make branch cuts (here at $x=0$) explicit.
In order to derive the desired permutation $t \leftrightarrow u$ we proceed by decomposing it into individual steps that only require the continuation of a single logarithm at a time. 
With this approach we are guaranteed that the argument of the logarithm in question corresponds to a ratio of Mandelstam variables.

We start in the scattering region introduced above where $s<0$ and $t<0$ and $x=\frac{t}{s}<1$.
\begin{enumerate}
\item We analytically continue to a region where $t > 0$ and introduce the new variable $x^\prime=-x$.
\bea
\log(x)&=&\log\left(\frac{t}{s}\right)\rightarrow \log\left(\frac{t}{-s}\right)- i \pi=\log(x^\prime)-i \pi.\nonumber\\
\log(1+x)&=&\log\left(\frac{u}{-s}\right)\rightarrow \log\left(\frac{u}{-s}\right)=\log\left(1-x^\prime\right).
\eea
\item Now we may exchange $t$ and $u$ without crossing any branch cut since both are positive.
\bea
\log(x^\prime)&\rightarrow&\log(1-x^{\prime\prime}).\nonumber\\
\log(1-x^\prime)&\rightarrow&\log(x^{\prime\prime}).
\eea
\item Finally, we analytically continue back to the region where our new $t<0$ and introduce the variable $x^{\prime\prime\prime}=-x^{\prime\prime}$.
\bea
\log(x^{\prime\prime})&=&\log\left(\frac{t}{-s}\right)\rightarrow \log\left(\frac{t}{s}\right)+ i \pi=\log(x^{\prime\prime\prime})+i \pi.\nonumber\\
\log(1-x^{\prime\prime})&=&\log\left(\frac{u}{-s}\right)\rightarrow \log\left(\frac{u}{-s}\right)=\log\left(1+x^{\prime\prime\prime}\right).
\eea
\end{enumerate}
Each of the above steps involves crossing only one branch cut with a well-defined $i\delta$ prescription. 

We then transform the remaining HPLs such that they have argument $x^\prime$ after step one, $x^{\prime\prime}$ after step two or $x^{\prime\prime\prime}$ after step three respectively. 
The required transformations  of the HPLs do not produce explicit imaginary parts. 
Finally, after step 3 we relabel $x^{\prime\prime\prime}\rightarrow x$ and finished our permutation procedure.
In summary we performed the transformation $x \rightarrow -1-x^{\prime\prime\prime}$.

Any other permutation of external legs can be found by a sequence of the two permutations described above. 
Notice, that the four graviton amplitude has to be fully symmetric under the exchange of any external momenta.
In order to ensure that we master the subtleties of the above procedure we computed all required master integrals in each permutation of external momenta separately (by solving for the boundary constants in each region) and find that they indeed can be transformed into one another by the above method.

We refer to the choice of variables $\{s,x\}$ for our amplitude as the s-channel representation since we scaled out all s dependence.
It may be convenient to represent a part of our functions differently  and we define the representations
\bea
\text{t-channel}: &&\left\{t,y=\frac{s}{t}\right\}.\nonumber\\
\text{u-channel}: &&\left\{u,z=\frac{u}{s}\right\}.
\eea
We eliminate the respectively missing Mandelstam invariant using the momentum conservation constraint.
It is interesting to ask how we can consistently perform a variable transformation on our amplitude given the difficulties with analytic continuation we examined above. Notice that the relations among the different variables
\beq
y=\frac{1}{x},\hspace{1cm}z=-1-x.
\eeq
correspond exactly to the transformation relations we considered in the case of our two basis permutations.
We can thus perform a variable transformation between our different representations by following the same steps as outlined above.
Note that changing the representation at the level of the amplitude amounts just to relabelling of $s$ and $x$ since the amplitude is already permutation invariant.

\section{The four-graviton amplitude up to three loops}
\label{sec:threeloops}

In this section we present the explicit results for the four-graviton amplitude up to third loop order. 
As mentioned earlier, the one- and two-loop results were already known but we recomputed them for the purpose of this work.
This serves as a useful cross check of our methods as well as for producing terms at higher order in the Laurent series in $\eps$ that are needed for the verifying the exponentiation of soft divergences.
The integrand for the one- and two-loop amplitude was presented in refs.~\cite{Bern:1998ug,Dunbar:1994bn,Green:1982sw}.

Integrands for the three loop amplitude were derived in refs.~\cite{Bern:2007hh,Bern:2008pv,Bern:2010ue} and we use in particular the representation given in ref.~\cite{Bern:2008pv}.
We perform a reduction of all appearing Feynman integrals to a basis of so-called master integrals by means of the Laporta algorithm~\cite{Laporta:2001dd} implemented in a private c++ code. 
The required master integrals for all amplitudes were previously calculated in order to derive the results of ref.~\cite{Henn:2016jdu}.
After inserting the expressions for the master integrals we obtain our final four graviton scattering amplitude.

The results obtained for the $L$ loop scattering amplitude $\mathcal{A}^{(L)}$ are of homogeneous transcendental weight $2L$. 
They depend on the ratio $x$ only via products harmonic poly-logarithms and low order polynomials of maximum rank $L$.
The harmonic polylogarithms only have indices 0 and $-1$.
The obtained functions $\mathcal{A}^{(L)}$ are the main result of this article and we present them in electronic form together with the arXiv submission of this article. 
In the following we also present explicit results for the finite coefficient functions of the logarithm of the scattering amplitude as defined in eq.~\eqref{eq:logamp}.

It was noted in ref.~\cite{BoucherVeronneau:2011qv} that the two loop amplitude could be written in particularly nice form by introducing a seed function $f^{(2)}(s,t,u)$ and summing over all permutations of the Mandelstam invariants.
\beq
F^{(2)}(s,t,u)=\sum\limits_{\sigma\in\text{Perms.}} \sigma(s) \sigma(t) f^{(2)}(\sigma(s),\sigma(t),\sigma(u)).
\eeq
From the above equation it is obvious to see that $F^{(2)}$ is invariant under the exchange of any two Mandelstam variables.
Note that the choice of $f$ is far from unique. 
For example we can consider the identity
\beq
0=\sum\limits_{\sigma\in\text{Perms.}} \sigma(s) \sigma(t) \log\left(\frac{\sigma(t)}{\sigma(s)}\right).
\eeq
Here we use a different choice of function $f^{(2)}$ compared to~\cite{BoucherVeronneau:2011qv}.
Our two-loop function is given by the pure function
\bea
\label{eq:2loopfunc}
f^{(2)}(s,t,u)&=& L_x H_{2,1}-H_{3,1}+\frac{L_x^4}{24}+\frac{1}{6} i \pi  L_x^3-\frac{1}{2} H_2 L_x^2\nonumber\\
&-&i \pi  H_2 L_x+H_3 L_x+i \pi  H_3-H_4+3 \zeta_4+i \pi  \zeta_3
\eea
Here, $L_x=\log x$. Furthermore, it can be written in terms of the integral representation
\beq
f^{(2)}(s,t,u)=3\zeta_4+i\pi\zeta_3+\int_0^x\frac{dx^\prime}{x^\prime} g^{(2)}(x^\prime) \,.
\eeq
with 
\beq
g^{(2)}(x)=\frac{\log ^3(x)}{6}-\frac{1}{2} \log (1+x) \log ^2(x)+\frac{1}{2} i \pi  \log ^2(x)+\frac{1}{2} \log ^2(1+x) \log (x)-i \pi  \log (1+x) \log (x).
\eeq
Note that the above integral has to be understood to be regulated at $x=0$ in the same way as introduced in eq.~\eqref{eq:reglogdef}.

A mathematical tool that has proven useful in the study of Feynman amplitudes and the functions appearing therein is the symbol map~\cite{Goncharov:1998kja}. 
In the case of HPLs the symbol map simply allows us to deduce the logarithmic integration kernels that define an HPL as introduced in eq.~\eqref{eq:HPLDef}.
For example,
\bea
\text{Symbol}(H_{0,0,1})&=&\text{Symbol}\left(\int_0^xd\log(x^\prime)\int_0^{x^\prime} d\log(x^{\prime\prime})\int_0^{x^{\prime\prime}} d\log(1-x^{\prime\prime\prime})\right)\nonumber\\
&=&\log(1-x) \circ \log(x)\circ\log(x).
\eea
The fact that the two-loop function takes the particular form of eq.~\eqref{eq:2loopfunc} allows us conclude that the last (right-most) entry of its symbol is given by $\log(x)$ since
\beq
d \,f^{(2)}(x) = d \log (x ) g^{(2)}(x).
\eeq
In order to express our three loop result we follow the inspiration derived from the two loop amplitude and write
\beq
F^{(3)}(s,t,u)=\sum\limits_{\sigma\in\text{Perms.}} \sigma(s^2) \sigma(t) f^{(3)}(\sigma(s),\sigma(t),\sigma(u)).
\eeq
We note that there is quite a degree of ambiguity in how to choose the function $f^{(3)}(s,t,u)$. We choose one particular definition with a relatively small number of terms.
\bea
&&f^{(3)}(s,t,u)=-\frac{1}{6} H_2 L_x^4-\frac{17}{21} i \pi  H_2 L_x^3+\frac{8}{3} H_3 L_x^3+\frac{719}{70} i \pi  H_3 L_x^2-\frac{41}{35} i \pi  H_2^2 L_x-2 H_3^2-\frac{93}{35} i \pi  H_2 H_3\nonumber\\ 
&&-\frac{5}{6} L_x^3 H_{2,1}-\frac{96}{35} i \pi  L_x^2 H_{2,1}-15 H_4 L_x^2-\frac{1384}{35} i \pi  H_4 L_x+40 H_5 L_x+\frac{419}{7} i \pi  H_5-40 H_6\nonumber\\ 
&&L_x^2 \left(-H_{3,1}\right)+H_2 L_x H_{2,1}+\frac{262}{35} i \pi  L_x H_{3,1}-2 H_{2,1}^2+4 i \pi  H_2 H_{2,1}-H_3 H_{2,1}+4 H_2 H_{3,1}\nonumber\\ 
&&4 L_x^2 H_{2,1,1}+4 L_x H_{3,2}+17 L_x H_{4,1}+\frac{87}{35} i \pi  L_x H_{2,1,1}+\frac{233}{35} i \pi  H_{3,2}+\frac{472}{35} i \pi  H_{4,1}-14 H_{5,1}\nonumber\\ 
&&-\frac{256}{35} i \pi  H_{2,2,1}-17 L_x H_{3,1,1}-\frac{739}{35} i \pi  H_{3,1,1}-4 H_{3,1,2}-11 H_{3,2,1}-11 H_{4,1,1}-6 L_x H_{2,1,1,1}\nonumber\\ 
&&8 H_{2,2,1,1}+38 H_{3,1,1,1}+\frac{79}{70} L_x^4 \zeta_2+\frac{179}{105} i \pi  L_x^3 \zeta_2-\frac{99}{35} H_2 L_x^2 \zeta_2-\frac{24}{35} i \pi  H_2 L_x \zeta_2+\frac{142}{35} H_2^2 \zeta_2\nonumber\\ 
&&-\frac{47}{5} L_x \zeta_2 H_{2,1}+i \pi  \zeta_2 H_{2,1}+\frac{132}{5} \zeta_2 H_{3,1}-\frac{47}{5} L_x H_3 \zeta_2-\frac{32}{5} i \pi  H_3 \zeta_2+\frac{132}{5} H_4 \zeta_2-\frac{149}{14} H_2 \zeta_4\nonumber\\ 
&&3 \zeta_3 H_{2,1}-8 i \pi  L_x^2 \zeta_3+\frac{23}{35} i \pi  H_2 \zeta_3-2 H_3 \zeta_3-\frac{3268}{35} i \pi  \zeta_2 \zeta_3-\frac{1615 \zeta_6}{8}-\frac{16 \zeta_3^2}{3}+\frac{75 i \pi  \zeta_5}{2}. 
\eea
Furthermore, we note that the function $f^{(3)}(s,t,u)$ may be written as 
\beq
f^{(3)}(s,t,u)=-\frac{1634}{105} i \pi ^3 \zeta (3)-\frac{16 \zeta (3)^2}{3}+\frac{75 i \pi  \zeta (5)}{2}-\frac{323 \pi ^6}{1512}+\int_0^x \frac{dx^\prime}{x^\prime} g^{(3)}(x^\prime).
\eeq
Note that this implies that we may choose a three loop function that equally has $\log(x)$ as the last entry of its symbol map.
Here, we introduced the function
\bea
g^{(3)}(x)=&&-\frac{1}{6} H_1 L_x^4-\frac{5}{12} H_1^2 L_x^3-\frac{17}{21} i \pi  H_1 L_x^3+\frac{2}{3} H_1^3 L_x^2-\frac{48}{35} i \pi  H_1^2 L_x^2-\frac{1}{4} H_1^4 L_x+\frac{29}{70} i \pi  H_1^3 L_x\nonumber\\ 
&&2 H_2 L_x^3+\frac{549}{70} i \pi  H_2 L_x^2+2 H_2^2 L_x+\frac{1}{2} H_1^2 H_2 L_x-\frac{82}{35} i \pi  H_1 H_2 L_x-\frac{1}{2} i \pi  H_2^2+2 i \pi  H_1^2 H_2\nonumber\\ 
&&-7 H_3 L_x^2-19 i \pi  H_3 L_x+10 H_4 L_x-\frac{1}{2} H_1^2 H_3-\frac{93}{35} i \pi  H_1 H_3-4 H_2 H_3+\frac{711}{35} i \pi  H_4\nonumber\\ 
&&-\frac{7}{2} L_x^2 H_{2,1}+H_1 L_x H_{2,1}+2 i \pi  L_x H_{2,1}+7 L_x H_{3,1}-2 H_1^2 H_{2,1}-\frac{116}{35} i \pi  H_1 H_{2,1}+\frac{268}{35} i \pi  H_{3,1}\nonumber\\ 
&&4 H_1 H_{3,1}+4 H_{3,2}+3 H_{4,1}-9 L_x H_{2,1,1}+8 H_1 H_{2,1,1}+\frac{116}{35} i \pi  H_{2,1,1}+H_{2,2,1}\nonumber\\ 
&&-4 H_{3,1,1}+\frac{158}{35} L_x^3 \zeta_2-\frac{99}{35} H_1 L_x^2 \zeta_2+\frac{179}{35} i \pi  L_x^2 \zeta_2-\frac{47}{10} H_1^2 L_x \zeta_2-\frac{24}{35} i \pi  H_1 L_x \zeta_2+\frac{1}{2} i \pi  H_1^2 \zeta_2\nonumber\\ 
&&17 \zeta_2 H_{2,1}-\frac{527}{35} H_2 L_x \zeta_2+\frac{284}{35} H_1 H_2 \zeta_2-\frac{248}{35} i \pi  H_2 \zeta_2+17 H_3 \zeta_2-\frac{149}{14} H_1 \zeta_4-16 i \pi  L_x \zeta_3\nonumber\\ 
&&\frac{3 H_1^2 \zeta_3}{2}+\frac{23}{35} i \pi  H_1 \zeta_3-2 H_2 \zeta_3.
\eea

The functions $F^{(n)}(x)$ are symmetric under permutation of Mandelstam invariants. 
To obtain them from the functions $f^{n}(x)$ the techniques for analytic continuation outlined in section~\ref{sec:analytic} can be used.
In ancillary files, we give explicit results in a computer-readable format.

\section{Soft and Regge limit}
\label{sec:regge}
Kinematic limits of scattering amplitudes often offer the possibility to explore universal properties of gauge theory. 
The limit of one particle having zero energy (soft limit)~\cite{BoucherVeronneau:2011qv,Bern:1998sv,Giddings:2009gj,Giddings:2010pp,Schnitzer:2007rn,Amati:1990xe} or scattering processes at very small angle (Regge or forward scatting limit)~\cite{Bartels:2012ra} are typical examples of this.
The fact that in such a limit the functional dependence of an amplitude becomes simpler and that scattering amplitudes factorise into universal building blocks and lower loop amplitudes can be exploited. 
Typically, logarithmically enhanced terms in such limits are of considerable interest.

Note that we write the tree-amplitude in superspace, hence it is completely symmetric under exchange of external particles.
For a discussion of the behavior of the tree amplitude for different helicity configurations of the scattered gravitons, see section VI. of  \cite{BoucherVeronneau:2011qv}.
Due to the large degree of symmetry of our four particle scattering amplitude many of these limits are related. 
The limit of forward scattering where $s$ becomes much greater than $t$ would send $x\rightarrow 0$. 
If we were to consider the soft limit where $s_{}\rightarrow 0$ we can simply obtain it be permuting the Mandelstam invariants $s\leftrightarrow t$. 
Since our amplitudes are invariant under such permutations we simply need to re-label the variables in order to obtain the desired result.

For convenience of the reader we display explicitly the logarithmically leading terms as $x\rightarrow 0$.
Note, that the leading logarithms are multiplied by powers of $t$.
At two loops we find
\bea
F^{(2)}_4  &=& -\frac{1}{12} t^2 \, \log^4(x)  \\
&+&\left[-\frac{1}{3} t u \log \left(-\frac{u}{s}\right)+\frac{1}{3} i \pi  t (s-t)  \right]\log^3(x) \nonumber\\
&+&\left[-t (2 s+t) H_{-2}-\frac{3}{2} \pi ^2 s t+i \pi  u^2 \log \left(-\frac{u}{s}\right)+\frac{1}{2} s u \log ^2\left(-\frac{u}{s}\right)\right]\log^2(x) +\dots \nonumber
\eea
Here, the dots indicate logarithmically suppressed terms. At three loops our result becomes
\bea
F^{(3)}_4  &=& \, \frac{7}{360} t^3 \, \log^6(x)  \\
&+&\left[\frac{7}{60} t^2 u \log \left(-\frac{u}{s}\right)-\frac{1}{20} i \pi  t^2 (3 s-2 t)  \right]\log^5(x) \nonumber\\
&+&\left[\frac{1}{6} t^2 (2 s+t) H_{-2}-\frac{1}{72} \pi ^2 t \left(6 s^2-75 s t+4 t^2\right)\right.\nonumber\\
&-&\left.\frac{1}{24} t u (s+5 t) \log ^2\left(-\frac{u}{s}\right)+\frac{1}{4} i \pi  t u (s+2 t) \log \left(-\frac{u}{s}\right) \right]\log^4(x) +\dots 
\eea
In ref. \cite{Bartels:2012ra}  all terms of the form 
\begin{align}
\left[ \frac{t}{s} \log^2 \frac{t}{s} \right]^L
\end{align}
were resummed to all orders, where $L$ is the loop order.
The resulting prediction provides a strong cross check and agrees with our formula.

The above formulae were organized as an expansion in terms of powers of logarithms.
The latter are in general power suppressed. 
In the strict limit $x \to 0$ we find 
\bea
\lim\limits_{x\rightarrow 0} F^{(2)}_4&=&3 \pi ^2 s^2 \zeta_{3} \epsilon +\mathcal{O}(x) +\mathcal{O}(\epsilon^2),\nonumber\\
\lim\limits_{x\rightarrow 0} F^{(3)}_4&=&\frac{2}{3} i \pi ^3 s^3 \zeta_{3} +\mathcal{O}(x) +\mathcal{O}(\epsilon) \,.
\eea
It would be interesting to understand these results from an analysis of the soft limit \cite{Bern:1998sv}.

\section{Conclusion and outlook}
\label{sec:conclusion}

We computed for the first time the three-loop four-graviton scattering amplitude in $\mathcal{N}=8$ supergravity.
Our result is written in closed form expressed in terms of harmonic polylogarithms of uniform weight six.
The amplitude has a complete permutation symmetry in the external legs.
We find that the amplitude can be written in an interesting way as the sum over permutations of a factor
$s t^2 f^{(3)}(x)$, where $f^{(3)}(x)$ satisfies a final-entry condition (in other words, it satisfies a simple differential equation). 
A similar property was found at two loops \cite{BoucherVeronneau:2011qv}.

It is fascinating that all amplitudes in $\mathcal{N}=8$ supergravity computed to date,
including the one computed in the present paper, have uniform and maximal transcendental weight,
just like their $\mathcal{N}=4$ sYM counterparts.
It would be very interesting to test this property for more general amplitudes.
Given the argument of \cite{Bern:2014kca}, the question whether the supergravity integrand
has double poles is tightly connected to the existence of BCJ-satisfying representations,
and to the behaviour at large loop momentum \cite{Bourjaily:2018omh}.
Reference \cite{Bourjaily:2018omh} reports that certain loop integrands have double and higher poles
at infinity, which would cast doubt on the uniform weight properties.
On the other hand, caveats may exist, and we find it interesting to ask whether supergravity amplitudes
in four dimensions turn out to be simpler than expected.
A hint in this direction is the recent work~\cite{Caron-Huot:2018ape} finding signs of integrability in N=8 supergravity.

Our result provides concise constraints on future calculations of three-loop five-graviton
amplitudes. In the soft and collinear limit, the latter are determined by our result.
We emphasise in particular that our result implies that any non-leading weight pieces in the three-loop
five-graviton amplitude has to vanish in the soft and collinear limit.

We expanded the answer in the Regge limit, and explicitly wrote out the leading power term,
as well as power suppressed terms. This allowed us to verify a prediction for certain leading logarithmic (but
power suppressed) terms made in ref. \cite{Bartels:2012ra}. 
It would be interesting if other terms could be predicted by Regge theory as well,
or if they could help understanding loop corrections to soft theorems \cite{Cachazo:2014fwa,Bern:2014oka,He:2014bga,DiVecchia:2018dob}.

\section{Acknowledgements}
It is a pleasure to thank J. Bourjaily, E. Herrmann, H. Johansson, and J. Trnka for enlightening discussions on loop integrands,
and Z. Bern for encouragement and helpful comments on the draft.
We thank the HPC group at JGU Mainz for support. 
This research received funding from the European Research Council (ERC) under the European Union's
Horizon 2020 research and innovation programme (grant agreement No 725110), {\it Novel structures in scattering amplitudes}.
B.~M. is supported by the Pappalardo fellowship.

\bibliography{refs2019}

\providecommand{\href}[2]{#2}\begingroup\raggedright\begin{thebibliography}{10}

\bibitem{Bern:2009kd}
Z.~Bern, J.~J. Carrasco, L.~J. Dixon, H.~Johansson, and R.~Roiban, {\it {The
  Ultraviolet Behavior of N=8 Supergravity at Four Loops}},  {\em Phys. Rev.
  Lett.} {\bf 103} (2009) 081301,
  [\href{http://xxx.lanl.gov/abs/0905.2326}{{\tt 0905.2326}}].

\bibitem{Bossard:2011tq}
G.~Bossard, P.~S. Howe, K.~S. Stelle, and P.~Vanhove, {\it {The vanishing
  volume of D=4 superspace}},  {\em Class. Quant. Grav.} {\bf 28} (2011)
  215005, [\href{http://xxx.lanl.gov/abs/1105.6087}{{\tt 1105.6087}}].

\bibitem{Bern:2018jmv}
Z.~Bern, J.~J. Carrasco, W.-M. Chen, A.~Edison, H.~Johansson,
  J.~Parra-Martinez, R.~Roiban, and M.~Zeng, {\it {Ultraviolet Properties of
  $\mathcal N = 8$ Supergravity at Five Loops}},  {\em Phys. Rev.} {\bf D98}
  (2018), no.~8 086021, [\href{http://xxx.lanl.gov/abs/1804.09311}{{\tt
  1804.09311}}].

\bibitem{Kawai:1985xq}
H.~Kawai, D.~C. Lewellen, and S.~H.~H. Tye, {\it {A Relation Between Tree
  Amplitudes of Closed and Open Strings}},  {\em Nucl. Phys.} {\bf B269} (1986)
  1--23.

\bibitem{Bern:1998ug}
Z.~Bern, L.~J. Dixon, D.~C. Dunbar, M.~Perelstein, and J.~S. Rozowsky, {\it {On
  the relationship between Yang-Mills theory and gravity and its implication
  for ultraviolet divergences}},  {\em Nucl. Phys.} {\bf B530} (1998) 401--456,
  [\href{http://xxx.lanl.gov/abs/hep-th/9802162}{{\tt hep-th/9802162}}].

\bibitem{Bern:2010ue}
Z.~Bern, J.~J.~M. Carrasco, and H.~Johansson, {\it {Perturbative Quantum
  Gravity as a Double Copy of Gauge Theory}},  {\em Phys. Rev. Lett.} {\bf 105}
  (2010) 061602, [\href{http://xxx.lanl.gov/abs/1004.0476}{{\tt 1004.0476}}].

\bibitem{Luna:2016hge}
A.~Luna, R.~Monteiro, I.~Nicholson, A.~Ochirov, D.~O'Connell, N.~Westerberg,
  and C.~D. White, {\it {Perturbative spacetimes from Yang-Mills theory}},
  {\em JHEP} {\bf 04} (2017) 069,
  [\href{http://xxx.lanl.gov/abs/1611.07508}{{\tt 1611.07508}}].

\bibitem{Luna:2016due}
A.~Luna, R.~Monteiro, I.~Nicholson, D.~O'Connell, and C.~D. White, {\it {The
  double copy: Bremsstrahlung and accelerating black holes}},  {\em JHEP} {\bf
  06} (2016) 023, [\href{http://xxx.lanl.gov/abs/1603.05737}{{\tt
  1603.05737}}].

\bibitem{Goldberger:2017vcg}
W.~D. Goldberger and A.~K. Ridgway, {\it {Bound states and the classical double
  copy}},  {\em Phys. Rev.} {\bf D97} (2018), no.~8 085019,
  [\href{http://xxx.lanl.gov/abs/1711.09493}{{\tt 1711.09493}}].

\bibitem{Luna:2017dtq}
A.~Luna, I.~Nicholson, D.~O'Connell, and C.~D. White, {\it {Inelastic Black
  Hole Scattering from Charged Scalar Amplitudes}},  {\em JHEP} {\bf 03} (2018)
  044, [\href{http://xxx.lanl.gov/abs/1711.03901}{{\tt 1711.03901}}].

\bibitem{Kosower:2018adc}
D.~A. Kosower, B.~Maybee, and D.~O'Connell, {\it {Amplitudes, Observables, and
  Classical Scattering}},  \href{http://xxx.lanl.gov/abs/1811.10950}{{\tt
  1811.10950}}.

\bibitem{Bern:2019nnu}
Z.~Bern, C.~Cheung, R.~Roiban, C.-H. Shen, M.~P. Solon, and M.~Zeng, {\it
  {Scattering Amplitudes and the Conservative Hamiltonian for Binary Systems at
  Third Post-Minkowskian Order}},
  \href{http://xxx.lanl.gov/abs/1901.04424}{{\tt 1901.04424}}.

\bibitem{Antonelli:2019ytb}
A.~Antonelli, A.~Buonanno, J.~Steinhoff, M.~van~de Meent, and J.~Vines, {\it
  {Energetics of two-body Hamiltonians in post-Minkowskian gravity}},
  \href{http://xxx.lanl.gov/abs/1901.07102}{{\tt 1901.07102}}.

\bibitem{Strominger:2017zoo}
A.~Strominger, {\it {Lectures on the Infrared Structure of Gravity and Gauge
  Theory}},  \href{http://xxx.lanl.gov/abs/1703.05448}{{\tt 1703.05448}}.

\bibitem{Cachazo:2014fwa}
F.~Cachazo and A.~Strominger, {\it {Evidence for a New Soft Graviton Theorem}},
   \href{http://xxx.lanl.gov/abs/1404.4091}{{\tt 1404.4091}}.

\bibitem{Bern:2014oka}
Z.~Bern, S.~Davies, and J.~Nohle, {\it {On Loop Corrections to Subleading Soft
  Behavior of Gluons and Gravitons}},  {\em Phys. Rev.} {\bf D90} (2014), no.~8
  085015, [\href{http://xxx.lanl.gov/abs/1405.1015}{{\tt 1405.1015}}].

\bibitem{He:2014bga}
S.~He, Y.-t. Huang, and C.~Wen, {\it {Loop Corrections to Soft Theorems in
  Gauge Theories and Gravity}},  {\em JHEP} {\bf 12} (2014) 115,
  [\href{http://xxx.lanl.gov/abs/1405.1410}{{\tt 1405.1410}}].

\bibitem{DiVecchia:2018dob}
P.~Vecchia, R.~Marotta, and M.~Mojaza, {\it {Multiloop Soft Theorem for
  Gravitons and Dilatons in the Bosonic String}},  {\em JHEP} {\bf 01} (2019)
  038, [\href{http://xxx.lanl.gov/abs/1808.04845}{{\tt 1808.04845}}].

\bibitem{Larkoski:2014bxa}
A.~J. Larkoski, D.~Neill, and I.~W. Stewart, {\it {Soft Theorems from Effective
  Field Theory}},  {\em JHEP} {\bf 06} (2015) 077,
  [\href{http://xxx.lanl.gov/abs/1412.3108}{{\tt 1412.3108}}].

\bibitem{Drummond:2006rz}
J.~M. Drummond, J.~Henn, V.~A. Smirnov, and E.~Sokatchev, {\it {Magic
  identities for conformal four-point integrals}},  {\em JHEP} {\bf 01} (2007)
  064, [\href{http://xxx.lanl.gov/abs/hep-th/0607160}{{\tt hep-th/0607160}}].

\bibitem{Alday:2007hr}
L.~F. Alday and J.~M. Maldacena, {\it {Gluon scattering amplitudes at strong
  coupling}},  {\em JHEP} {\bf 06} (2007) 064,
  [\href{http://xxx.lanl.gov/abs/0705.0303}{{\tt 0705.0303}}].

\bibitem{Drummond:2007au}
J.~M. Drummond, J.~Henn, G.~P. Korchemsky, and E.~Sokatchev, {\it {Conformal
  Ward identities for Wilson loops and a test of the duality with gluon
  amplitudes}},  {\em Nucl. Phys.} {\bf B826} (2010) 337--364,
  [\href{http://xxx.lanl.gov/abs/0712.1223}{{\tt 0712.1223}}].

\bibitem{Drummond:2008vq}
J.~M. Drummond, J.~Henn, G.~P. Korchemsky, and E.~Sokatchev, {\it {Dual
  superconformal symmetry of scattering amplitudes in N=4 super-Yang-Mills
  theory}},  {\em Nucl. Phys.} {\bf B828} (2010) 317--374,
  [\href{http://xxx.lanl.gov/abs/0807.1095}{{\tt 0807.1095}}].

\bibitem{Berkovits:2008ic}
N.~Berkovits and J.~Maldacena, {\it {Fermionic T-Duality, Dual Superconformal
  Symmetry, and the Amplitude/Wilson Loop Connection}},  {\em JHEP} {\bf 09}
  (2008) 062, [\href{http://xxx.lanl.gov/abs/0807.3196}{{\tt 0807.3196}}].

\bibitem{Brandhuber:2008pf}
A.~Brandhuber, P.~Heslop, and G.~Travaglini, {\it {A Note on dual
  superconformal symmetry of the N=4 super Yang-Mills S-matrix}},  {\em Phys.
  Rev.} {\bf D78} (2008) 125005, [\href{http://xxx.lanl.gov/abs/0807.4097}{{\tt
  0807.4097}}].

\bibitem{Drummond:2009fd}
J.~M. Drummond, J.~M. Henn, and J.~Plefka, {\it {Yangian symmetry of scattering
  amplitudes in N=4 super Yang-Mills theory}},  {\em JHEP} {\bf 05} (2009) 046,
  [\href{http://xxx.lanl.gov/abs/0902.2987}{{\tt 0902.2987}}].

\bibitem{Caron-Huot:2014gia}
S.~Caron-Huot and J.~M. Henn, {\it {Solvable Relativistic Hydrogenlike System
  in Supersymmetric Yang-Mills Theory}},  {\em Phys. Rev. Lett.} {\bf 113}
  (2014), no.~16 161601, [\href{http://xxx.lanl.gov/abs/1408.0296}{{\tt
  1408.0296}}].

\bibitem{Bern:2017gdk}
Z.~Bern, M.~Enciso, H.~Ita, and M.~Zeng, {\it {Dual Conformal Symmetry,
  Integration-by-Parts Reduction, Differential Equations and the Nonplanar
  Sector}},  {\em Phys. Rev.} {\bf D96} (2017), no.~9 096017,
  [\href{http://xxx.lanl.gov/abs/1709.06055}{{\tt 1709.06055}}].

\bibitem{Ben-Israel:2018ckc}
R.~Ben-Israel, A.~G. Tumanov, and A.~Sever, {\it {Scattering Amplitudes --
  Wilson Loops Duality for the First Non-planar Correction}},
  \href{http://xxx.lanl.gov/abs/1802.09395}{{\tt 1802.09395}}.

\bibitem{Bern:2018oao}
Z.~Bern, M.~Enciso, C.-H. Shen, and M.~Zeng, {\it {Dual Conformal Structure
  Beyond the Planar Limit}},  \href{http://xxx.lanl.gov/abs/1806.06509}{{\tt
  1806.06509}}.

\bibitem{Chicherin:2018wes}
D.~Chicherin, J.~M. Henn, and E.~Sokatchev, {\it {Implications of nonplanar
  dual conformal symmetry}},  {\em JHEP} {\bf 09} (2018) 012,
  [\href{http://xxx.lanl.gov/abs/1807.06321}{{\tt 1807.06321}}].

\bibitem{Bianchi:2018rrj}
L.~Bianchi, A.~Brandhuber, R.~Panerai, and G.~Travaglini, {\it {Dual conformal
  invariance for form factors}},
  \href{http://xxx.lanl.gov/abs/1812.10468}{{\tt 1812.10468}}.

\bibitem{Caron-Huot:2018ape}
S.~Caron-Huot and Z.~Zahraee, {\it {Integrability of Black Hole Orbits in
  Maximal Supergravity}},  \href{http://xxx.lanl.gov/abs/1810.04694}{{\tt
  1810.04694}}.

\bibitem{Kotikov:2004er}
A.~V. Kotikov, L.~N. Lipatov, A.~I. Onishchenko, and V.~N. Velizhanin, {\it
  {Three loop universal anomalous dimension of the Wilson operators in $N=4$
  SUSY Yang-Mills model}},  {\em Phys. Lett.} {\bf B595} (2004) 521--529,
  [\href{http://xxx.lanl.gov/abs/hep-th/0404092}{{\tt hep-th/0404092}}].
  [Erratum: Phys. Lett.B632,754(2006)].

\bibitem{KOTIKOV2007217}
A.~Kotikov and L.~Lipatov, {\it On the highest transcendentality in n=4 susy},
  {\em Nuclear Physics B} {\bf 769} (2007), no.~3 217 -- 255.

\bibitem{ArkaniHamed:2010gh}
N.~Arkani-Hamed, J.~L. Bourjaily, F.~Cachazo, and J.~Trnka, {\it {Local
  Integrals for Planar Scattering Amplitudes}},  {\em JHEP} {\bf 06} (2012)
  125, [\href{http://xxx.lanl.gov/abs/1012.6032}{{\tt 1012.6032}}].

\bibitem{Henn:2013pwa}
J.~M. Henn, {\it {Multiloop integrals in dimensional regularization made
  simple}},  {\em Phys. Rev. Lett.} {\bf 110} (2013) 251601,
  [\href{http://xxx.lanl.gov/abs/1304.1806}{{\tt 1304.1806}}].

\bibitem{Arkani-Hamed:2014via}
N.~Arkani-Hamed, J.~L. Bourjaily, F.~Cachazo, and J.~Trnka, {\it {Singularity
  Structure of Maximally Supersymmetric Scattering Amplitudes}},  {\em Phys.
  Rev. Lett.} {\bf 113} (2014), no.~26 261603,
  [\href{http://xxx.lanl.gov/abs/1410.0354}{{\tt 1410.0354}}].

\bibitem{WasserMSc}
P.~Wasser, {\it {Analytic properties of Feynman integrals for scattering
  amplitudes}},  {\em MSc.} (2016)
  [\href{http://xxx.lanl.gov/abs/https://publications.ub.uni-mainz.de/theses/frontdoor.php?source
  opus=100001967}{{\tt
  https://publications.ub.uni-mainz.de/theses/frontdoor.php?source
  opus=100001967}}].

\bibitem{Frellesvig:2017aai}
H.~Frellesvig and C.~G. Papadopoulos, {\it {Cuts of Feynman Integrals in Baikov
  representation}},  {\em JHEP} {\bf 04} (2017) 083,
  [\href{http://xxx.lanl.gov/abs/1701.07356}{{\tt 1701.07356}}].

\bibitem{Harley:2017qut}
M.~Harley, F.~Moriello, and R.~M. Schabinger, {\it {Baikov-Lee Representations
  Of Cut Feynman Integrals}},  {\em JHEP} {\bf 06} (2017) 049,
  [\href{http://xxx.lanl.gov/abs/1705.03478}{{\tt 1705.03478}}].

\bibitem{Primo:2016ebd}
A.~Primo and L.~Tancredi, {\it {On the maximal cut of Feynman integrals and the
  solution of their differential equations}},  {\em Nucl. Phys.} {\bf B916}
  (2017) 94--116, [\href{http://xxx.lanl.gov/abs/1610.08397}{{\tt
  1610.08397}}].

\bibitem{Chicherin:2018old}
D.~Chicherin, T.~Gehrmann, J.~M. Henn, P.~Wasser, Y.~Zhang, and S.~Zoia, {\it
  {All master integrals for three-jet production at NNLO}},
  \href{http://xxx.lanl.gov/abs/1812.11160}{{\tt 1812.11160}}.

\bibitem{Chicherin:2019xeg}
D.~Chicherin, T.~Gehrmann, J.~M. Henn, P.~Wasser, Y.~Zhang, and S.~Zoia, {\it
  {The two-loop five-particle amplitude in $\mathcal{N}=8$ supergravity}},
  \href{http://xxx.lanl.gov/abs/1901.05932}{{\tt 1901.05932}}.

\bibitem{Abreu:2019rpt}
S.~Abreu, L.~J. Dixon, E.~Herrmann, B.~Page, and M.~Zeng, {\it {The two-loop
  five-point amplitude in $\mathcal N=8$ supergravity}},
  \href{http://xxx.lanl.gov/abs/1901.08563}{{\tt 1901.08563}}.

\bibitem{Naculich:2008ew}
S.~G. Naculich, H.~Nastase, and H.~J. Schnitzer, {\it {Two-loop graviton
  scattering relation and IR behavior in N=8 supergravity}},  {\em Nucl. Phys.}
  {\bf B805} (2008) 40--58, [\href{http://xxx.lanl.gov/abs/0805.2347}{{\tt
  0805.2347}}].

\bibitem{Brandhuber:2008tf}
A.~Brandhuber, P.~Heslop, A.~Nasti, B.~Spence, and G.~Travaglini, {\it
  {Four-point Amplitudes in N=8 Supergravity and Wilson Loops}},  {\em Nucl.
  Phys.} {\bf B807} (2009) 290--314,
  [\href{http://xxx.lanl.gov/abs/0805.2763}{{\tt 0805.2763}}].

\bibitem{BoucherVeronneau:2011qv}
C.~Boucher-Veronneau and L.~J. Dixon, {\it {N $\ge$ 4 Supergravity Amplitudes
  from Gauge Theory at Two Loops}},  {\em JHEP} {\bf 12} (2011) 046,
  [\href{http://xxx.lanl.gov/abs/1110.1132}{{\tt 1110.1132}}].

\bibitem{Bern:2007hh}
Z.~Bern, J.~J. Carrasco, L.~J. Dixon, H.~Johansson, D.~A. Kosower, and
  R.~Roiban, {\it {Three-Loop Superfiniteness of N=8 Supergravity}},  {\em
  Phys. Rev. Lett.} {\bf 98} (2007) 161303,
  [\href{http://xxx.lanl.gov/abs/hep-th/0702112}{{\tt hep-th/0702112}}].

\bibitem{Bern:2008pv}
Z.~Bern, J.~J.~M. Carrasco, L.~J. Dixon, H.~Johansson, and R.~Roiban, {\it
  {Manifest Ultraviolet Behavior for the Three-Loop Four-Point Amplitude of N=8
  Supergravity}},  {\em Phys. Rev.} {\bf D78} (2008) 105019,
  [\href{http://xxx.lanl.gov/abs/0808.4112}{{\tt 0808.4112}}].

\bibitem{Bern:2008qj}
Z.~Bern, J.~J.~M. Carrasco, and H.~Johansson, {\it {New Relations for
  Gauge-Theory Amplitudes}},  {\em Phys. Rev.} {\bf D78} (2008) 085011,
  [\href{http://xxx.lanl.gov/abs/0805.3993}{{\tt 0805.3993}}].

\bibitem{Bern:2014kca}
Z.~Bern, E.~Herrmann, S.~Litsey, J.~Stankowicz, and J.~Trnka, {\it {Logarithmic
  Singularities and Maximally Supersymmetric Amplitudes}},  {\em JHEP} {\bf 06}
  (2015) 202, [\href{http://xxx.lanl.gov/abs/1412.8584}{{\tt 1412.8584}}].

\bibitem{Bourjaily:2018omh}
J.~L. Bourjaily, E.~Herrmann, and J.~Trnka, {\it {Amplitudes at Infinity}},
  \href{http://xxx.lanl.gov/abs/1812.11185}{{\tt 1812.11185}}.

\bibitem{Henn:2016jdu}
J.~M. Henn and B.~Mistlberger, {\it {Four-Gluon Scattering at Three Loops,
  Infrared Structure, and the Regge Limit}},  {\em Phys. Rev. Lett.} {\bf 117}
  (2016), no.~17 171601, [\href{http://xxx.lanl.gov/abs/1608.00850}{{\tt
  1608.00850}}].

\bibitem{Remiddi:1999ew}
E.~Remiddi and J.~A.~M. Vermaseren, {\it {Harmonic polylogarithms}},  {\em Int.
  J. Mod. Phys.} {\bf A15} (2000) 725--754,
  [\href{http://xxx.lanl.gov/abs/hep-ph/9905237}{{\tt hep-ph/9905237}}].

\bibitem{Bartels:2012ra}
J.~Bartels, L.~N. Lipatov, and A.~Sabio~Vera, {\it {Double-logarithms in
  Einstein-Hilbert gravity and supergravity}},  {\em JHEP} {\bf 07} (2014) 056,
  [\href{http://xxx.lanl.gov/abs/1208.3423}{{\tt 1208.3423}}].

\bibitem{ArkaniHamed:2008gz}
N.~Arkani-Hamed, F.~Cachazo, and J.~Kaplan, {\it {What is the Simplest Quantum
  Field Theory?}},  {\em JHEP} {\bf 09} (2010) 016,
  [\href{http://xxx.lanl.gov/abs/0808.1446}{{\tt 0808.1446}}].

\bibitem{Goncharov:1998kja}
A.~B. Goncharov, {\it {Multiple polylogarithms, cyclotomy and modular
  complexes}},  {\em Math. Res. Lett.} {\bf 5} (1998) 497--516,
  [\href{http://xxx.lanl.gov/abs/1105.2076}{{\tt 1105.2076}}].

\bibitem{Panzer:2014gra}
E.~Panzer, {\it {On hyperlogarithms and Feynman integrals with divergences and
  many scales}},  {\em JHEP} {\bf 03} (2014) 071,
  [\href{http://xxx.lanl.gov/abs/1401.4361}{{\tt 1401.4361}}].

\bibitem{Duhr:2011zq}
C.~Duhr, H.~Gangl, and J.~R. Rhodes, {\it {From polygons and symbols to
  polylogarithmic functions}},  {\em JHEP} {\bf 10} (2012) 075,
  [\href{http://xxx.lanl.gov/abs/1110.0458}{{\tt 1110.0458}}].

\bibitem{Panzer:2014caa}
E.~Panzer, {\it {Algorithms for the symbolic integration of hyperlogarithms
  with applications to Feynman integrals}},  {\em Comput. Phys. Commun.} {\bf
  188} (2015) 148--166, [\href{http://xxx.lanl.gov/abs/1403.3385}{{\tt
  1403.3385}}].

\bibitem{Duhr:2012fh}
C.~Duhr, {\it {Hopf algebras, coproducts and symbols: an application to Higgs
  boson amplitudes}},  {\em JHEP} {\bf 08} (2012) 043,
  [\href{http://xxx.lanl.gov/abs/1203.0454}{{\tt 1203.0454}}].

\bibitem{Maitre:2005uu}
D.~Maitre, {\it {HPL, a mathematica implementation of the harmonic
  polylogarithms}},  {\em Comput. Phys. Commun.} {\bf 174} (2006) 222--240,
  [\href{http://xxx.lanl.gov/abs/hep-ph/0507152}{{\tt hep-ph/0507152}}].

\bibitem{Weinberg:1965nx}
S.~Weinberg, {\it {Infrared photons and gravitons}},  {\em Phys. Rev.} {\bf
  140} (1965) B516--B524.

\bibitem{Akhoury:2011kq}
R.~Akhoury, R.~Saotome, and G.~Sterman, {\it {Collinear and Soft Divergences in
  Perturbative Quantum Gravity}},  {\em Phys. Rev.} {\bf D84} (2011) 104040,
  [\href{http://xxx.lanl.gov/abs/1109.0270}{{\tt 1109.0270}}].

\bibitem{Beneke:2012xa}
M.~Beneke and G.~Kirilin, {\it {Soft-collinear gravity}},  {\em JHEP} {\bf 09}
  (2012) 066, [\href{http://xxx.lanl.gov/abs/1207.4926}{{\tt 1207.4926}}].

\bibitem{VanNieuwenhuizen:1973qf}
P.~Van~Nieuwenhuizen, {\it {Radiation of massive gravitation}},  {\em Phys.
  Rev.} {\bf D7} (1973) 2300--2308.

\bibitem{Naculich:2011ry}
S.~G. Naculich and H.~J. Schnitzer, {\it {Eikonal methods applied to
  gravitational scattering amplitudes}},  {\em JHEP} {\bf 05} (2011) 087,
  [\href{http://xxx.lanl.gov/abs/1101.1524}{{\tt 1101.1524}}].

\bibitem{White:2011yy}
C.~D. White, {\it {Factorization Properties of Soft Graviton Amplitudes}},
  {\em JHEP} {\bf 05} (2011) 060,
  [\href{http://xxx.lanl.gov/abs/1103.2981}{{\tt 1103.2981}}].

\bibitem{Almelid:2015jia}
O.~Almelid, C.~Duhr, and E.~Gardi, {\it {Three-loop corrections to the soft
  anomalous dimension in multileg scattering}},  {\em Phys. Rev. Lett.} {\bf
  117} (2016), no.~17 172002, [\href{http://xxx.lanl.gov/abs/1507.00047}{{\tt
  1507.00047}}].

\bibitem{Almelid:2017qju}
O.~Almelid, C.~Duhr, E.~Gardi, A.~McLeod, and C.~D. White, {\it {Bootstrapping
  the QCD soft anomalous dimension}},  {\em JHEP} {\bf 09} (2017) 073,
  [\href{http://xxx.lanl.gov/abs/1706.10162}{{\tt 1706.10162}}].

\bibitem{White:2015wha}
C.~D. White, {\it {An Introduction to Webs}},  {\em J. Phys.} {\bf G43} (2016),
  no.~3 033002, [\href{http://xxx.lanl.gov/abs/1507.02167}{{\tt 1507.02167}}].

\bibitem{Bern:2017rjw}
Z.~Bern, J.~Parra-Martinez, and R.~Roiban, {\it {Canceling the U(1) Anomaly in
  the $S$ Matrix of $N$=4 Supergravity}},  {\em Phys. Rev. Lett.} {\bf 121}
  (2018), no.~10 101604, [\href{http://xxx.lanl.gov/abs/1712.03928}{{\tt
  1712.03928}}].

\bibitem{Bern:2017tuc}
Z.~Bern, A.~Edison, D.~Kosower, and J.~Parra-Martinez, {\it {Curvature-squared
  multiplets, evanescent effects, and the U(1) anomaly in $N=4$ supergravity}},
   {\em Phys. Rev.} {\bf D96} (2017), no.~6 066004,
  [\href{http://xxx.lanl.gov/abs/1706.01486}{{\tt 1706.01486}}].

\bibitem{Bern:2013uka}
Z.~Bern, S.~Davies, T.~Dennen, A.~V. Smirnov, and V.~A. Smirnov, {\it
  {Ultraviolet Properties of N=4 Supergravity at Four Loops}},  {\em Phys. Rev.
  Lett.} {\bf 111} (2013), no.~23 231302,
  [\href{http://xxx.lanl.gov/abs/1309.2498}{{\tt 1309.2498}}].

\bibitem{Carrasco:2013ypa}
J.~J.~M. Carrasco, R.~Kallosh, R.~Roiban, and A.~A. Tseytlin, {\it {On the U(1)
  duality anomaly and the S-matrix of N=4 supergravity}},  {\em JHEP} {\bf 07}
  (2013) 029, [\href{http://xxx.lanl.gov/abs/1303.6219}{{\tt 1303.6219}}].

\bibitem{Freedman:2017zgq}
D.~Z. Freedman, R.~Kallosh, D.~Murli, A.~Van~Proeyen, and Y.~Yamada, {\it
  {Absence of U(1) Anomalous Superamplitudes in $\mathcal{N}\geq 5$
  Supergravities}},  {\em JHEP} {\bf 05} (2017) 067,
  [\href{http://xxx.lanl.gov/abs/1703.03879}{{\tt 1703.03879}}].

\bibitem{Kallosh:2016xnm}
R.~Kallosh, {\it {Cancellation of Conformal and Chiral Anomalies in
  $\mathcal{N} \geq 5$ supergravities}},  {\em Phys. Rev.} {\bf D95} (2017),
  no.~4 041701, [\href{http://xxx.lanl.gov/abs/1612.08978}{{\tt 1612.08978}}].

\bibitem{Tausk:1999vh}
J.~B. Tausk, {\it {Nonplanar massless two loop Feynman diagrams with four
  on-shell legs}},  {\em Phys. Lett.} {\bf B469} (1999) 225--234,
  [\href{http://xxx.lanl.gov/abs/hep-ph/9909506}{{\tt hep-ph/9909506}}].

\bibitem{Henn:2013nsa}
J.~M. Henn, A.~V. Smirnov, and V.~A. Smirnov, {\it {Evaluating single-scale
  and/or non-planar diagrams by differential equations}},  {\em JHEP} {\bf 03}
  (2014) 088, [\href{http://xxx.lanl.gov/abs/1312.2588}{{\tt 1312.2588}}].

\bibitem{Dunbar:1994bn}
D.~C. Dunbar and P.~S. Norridge, {\it {Calculation of graviton scattering
  amplitudes using string based methods}},  {\em Nucl. Phys.} {\bf B433} (1995)
  181--208, [\href{http://xxx.lanl.gov/abs/hep-th/9408014}{{\tt
  hep-th/9408014}}].

\bibitem{Green:1982sw}
M.~B. Green, J.~H. Schwarz, and L.~Brink, {\it {N=4 Yang-Mills and N=8
  Supergravity as Limits of String Theories}},  {\em Nucl. Phys.} {\bf B198}
  (1982) 474--492.

\bibitem{Laporta:2001dd}
S.~Laporta, {\it {High precision calculation of multiloop Feynman integrals by
  difference equations}},  {\em Int. J. Mod. Phys.} {\bf A15} (2000)
  5087--5159, [\href{http://xxx.lanl.gov/abs/hep-ph/0102033}{{\tt
  hep-ph/0102033}}].

\bibitem{Bern:1998sv}
Z.~Bern, L.~J. Dixon, M.~Perelstein, and J.~S. Rozowsky, {\it {Multileg one
  loop gravity amplitudes from gauge theory}},  {\em Nucl. Phys.} {\bf B546}
  (1999) 423--479, [\href{http://xxx.lanl.gov/abs/hep-th/9811140}{{\tt
  hep-th/9811140}}].

\bibitem{Giddings:2009gj}
S.~B. Giddings and R.~A. Porto, {\it {The Gravitational S-matrix}},  {\em Phys.
  Rev.} {\bf D81} (2010) 025002, [\href{http://xxx.lanl.gov/abs/0908.0004}{{\tt
  0908.0004}}].

\bibitem{Giddings:2010pp}
S.~B. Giddings, M.~Schmidt-Sommerfeld, and J.~R. Andersen, {\it {High energy
  scattering in gravity and supergravity}},  {\em Phys. Rev.} {\bf D82} (2010)
  104022, [\href{http://xxx.lanl.gov/abs/1005.5408}{{\tt 1005.5408}}].

\bibitem{Schnitzer:2007rn}
H.~J. Schnitzer, {\it {Reggeization of N=8 supergravity and N=4 Yang-Mills
  theory. II.}},  \href{http://xxx.lanl.gov/abs/0706.0917}{{\tt 0706.0917}}.

\bibitem{Amati:1990xe}
D.~Amati, M.~Ciafaloni, and G.~Veneziano, {\it {Higher Order Gravitational
  Deflection and Soft Bremsstrahlung in Planckian Energy Superstring
  Collisions}},  {\em Nucl. Phys.} {\bf B347} (1990) 550--580.

\end{thebibliography}\endgroup
\bibliographystyle{JHEP}

\end{document}